\documentclass[10pt,twocolumn]{article}
\usepackage[T1]{fontenc}
\usepackage[utf8]{inputenc}
\pdfoutput=1
\usepackage{graphicx}
\usepackage{fancyvrb}
\usepackage{hyperref}
\usepackage{authblk}

\title{Towards a Polyglot Data Access Layer for a Low-Code Application Development Platform\thanks{This work was supported by Lisboa2020, Compete2020 and FEDER through Project RADicalize (LISBOA-01-0247-FEDER-017116 | POCI-01-0247-FEDER-017116).}}

\author[1]{Ana Nunes Alonso}
\author[2]{Jo\~{a}o Abreu}
\author[2]{David Nunes}
\author[2]{Andr\'{e} Vieira} 
\author[2]{\and Luiz Santos}
\author[2]{T\'{e}rcio Soares}
\author[1]{Jos\'{e} Pereira}

\affil[1]{INESC TEC and U. Minho\\\texttt{ana.n.alonso@inesctec.pt}, \texttt{jop@di.uminho.pt}}
\affil[2]{OutSystems\\\texttt{\textit{first.last}@outsystems.com}}
 
 \date{}
 
\begin{document}

\maketitle

\begin{abstract}
Low-code application development as proposed by the OutSystems Platform enables fast mobile and desktop application development and deployment. It hinges on visual development of the interface and business logic but also on easy integration with data stores and services while delivering robust applications that scale.

Data integration increasingly means accessing a variety of NoSQL stores. Unfortunately, the diversity of data and processing models, that make them useful in the first place, is difficult to reconcile with the simplification of abstractions exposed to developers in a low-code platform. Moreover, NoSQL data stores also rely on a variety of general purpose and custom scripting languages as their main interfaces.

In this paper we propose a polyglot data access layer for the OutSystems Platform that uses SQL with optional embedded script snippets to bridge the gap between low-code and full access to NoSQL stores. 
In detail, we characterize the challenges for integrating a variety of NoSQL data stores; we describe the architecture and proof-of-concept implementation; and evaluate it with a sample application.
\end{abstract}

\section{Introduction}
\label{sec:intro}
According to Forrester Research, that defines low-code as ``enabl[ing] rapid delivery of business applications with a minimum of hand-coding and minimal upfront investment in setup, training, and deployment,'' OutSystems is a leading low-code platform for application development and delivery\,\cite{forrester}. It emphasizes drag-and-drop to define the functionality for UI, business processes, logic, and data models to create full-stack, cross-platform applications.

Integrating with existing systems increasingly means connecting to a variety of NoSQL data stores, deployed as businesses take advantage of Big Data. Our goal is to enable interactive applications built with the OutSystems Platform \cite{outsystems:arch} to query data in NoSQL stores.

The current standard for integrating NoSQL stores with available low-code platforms is for developers to manually define how the available data must be imported and consumed by the platform, requiring expertise in each particular NoSQL store, especially if performance is a concern.
 
Conversely, an ideal OutSystems experience for leveraging Big Data should be:
(1) create an integration with the Big Data repository, providing the connection details
including credentials; 
(2) the platform introspects the data in the repository and creates a representation for it;
(3) the developer trims down the mapping to the information that is valuable for the application;
(4) the developer is able to query this information with some transformation capabilities which include filtering, sorting, grouping;
(5) the platform handles all the requirements for providing this information with enterprise-grade non-functional requirements (NFRs) such as security, performance, and scalability; and 
(6) the developer is able to create applications that can interact with the Big Data repository and leverage this information in business logic and processes, and also to create visualizations.

Delivering the ideal experience for Big Data does however raise significant challenges. First, step (2) is challenged by NoSQL systems not having a standardized data model, a standard method to query metadata, or even in many cases by not enforcing a schema at all. In fact, even if stored data conform to a well defined schema, it may be only implicit in the code of applications that manipulate it.

Second, the value added by NoSQL data stores rests precisely on a diversity of query operations and query composition mechanisms, that exploit specific data models, storage, and indexing structures. Exposing these as visual abstractions for manipulation in step (4) risks polluting the low-code platform with multiple particular and overlapping concepts, instead of general purpose abstractions. On the other hand, if we expose the minimal common factor between all NoSQL data stores, we are likely to end up with minimal filtering capabilities that prevent developers from fully exploiting NoSQL integration. In either case, some NoSQL data stores offer only very minimal query processing capabilities and thus force client applications to code all other data manipulation operations, which also conflicts with the low-code approach.
Finally, step (5) requires ensuring that performance is compatible with interactive applications means that one cannot resort to built-in MapReduce to cope with missing query functionality, as it leads to high latency and resource usage. Also, coping with large scale data sets means full data traversals should be avoided. This can be done by exposing relevant indexing mechanisms and resorting to approximate and incomplete data, for instance, when displaying a developer preview during step (3).

Our goal is to add the ability to query data in NoSQL stores to interactive applications creating using the OutSystems platform, preserving the low-code developer experience, i.e. without requiring specific NoSQL knowledge for a developer to successfully write applications that leverage this type of data stores.

Enabling the seamless integration of a multitude NoSQL stores with the OutSystems platform will offer its more than 200 000 developers a considerable competitive advantage over other currently available low-code offers. Moreover, Gartner predicts low code application platforms will be used for 65\% of all application development activity in 5 years time\cite{gartner}, amplifying the impact of our contribution.

In this paper we summarize our work on a proof-of-concept polyglot data access layer for the OutSystems Platform that addresses these challenges, thus making the following contributions:
\begin{itemize}
\item We describe in detail the challenge in integrating NoSQL data stores in a low-code development platform targeted at relational data. This is achieved mainly by surveying how data and query models in a spectrum of NoSQL data stores match the abstractions that underlie the low-code approach in OutSystems.
\item We propose to use a polyglot query engine, based on extended relational data and query models, with embedded NoSQL query script fragments as the approach that reconciles the expectation of low-code integration with the reality of NoSQL diversity.
\item We describe a proof-of-concept implementation that leverages an off-the-shelf SQL query engine that implements the SQL/MED standard \cite{sqlmed} for managing external data. 
\end{itemize}
~\\[-18pt]
As a result, we describe various lessons learned, that are relevant to the integration of NoSQL data stores with low-code tools in general, to how NoSQL data stores can evolve to make this integration easier and more effective, and to research and development in polyglot query processing systems in general.

The rest of the paper is structured as follows.  
In Section~\ref{sec:bg} we briefly describe the OutSystems platform, how it handles data access, and integrates with external SQL databases. Section~\ref{sec:nosql} presents the main results of an analysis of target data stores focusing on how their characteristics impact our goal.  Section~\ref{sec:prop} describes our proposal to integrate NoSQL data stores in the OutSystems platform, including our current proof-of-concept implementation. This proposal is then evaluated in Section~\ref{sec:eval} and compared to related work in Section~\ref{sec:related}. Section~\ref{sec:concl} concludes the paper by discussing the main lessons learned.
\section{Background}
\label{sec:bg}

The OutSystems Platform is used to develop, deploy and operate a large number of custom
applications through a model driven approach and based on a low code visual language \cite{outsystems:arch}.
A key trait of the approach is that 
the most common tasks executed in the creation of an
Information System are visually modeled. This is a major contribution for
providing acceleration to customers. Visual models abstract a lot
of the complexity of low level coding, are easier to read and therefore reduce the time
needed for knowledge transfer. Having a common modeling language favors skill reuse
either when implementing back end logic, business processes, web UIs, or mobile UIs.
A second key trait is that customers should never hit a wall, in the sense that the not so
common tasks can also be achieved, while sometimes not as easily, but still using visual
modelling and/or extensibility to 3GL languages.

\subsection{Architecture}

\begin{figure}[t]
 \centering
 \includegraphics[width=1\columnwidth]{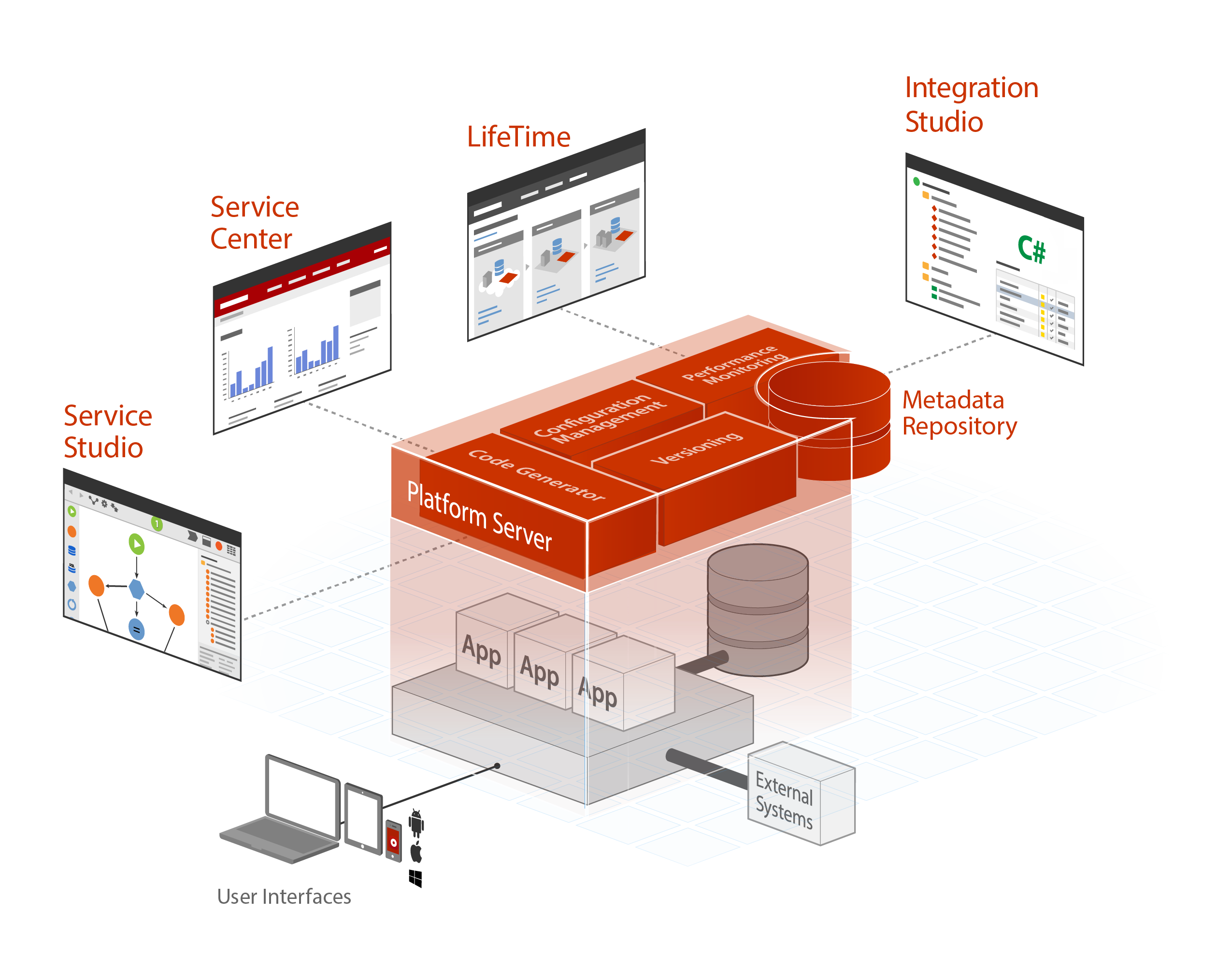}.
 \caption{OutSystems Platform Architecture}
 \label{fig:outsystems}
\end{figure}

The OutSystems solution architecture\,\cite{outsystems:platform} features five main components, as shown in Figure~\ref{fig:outsystems}:

\textbf{Service Studio.} The development environment for all the DSLs supported by OutSystems. When the developer publishes an application, Service Studio saves a document with the application's model and sends it to the Platform Server. 

\textbf{Platform Server.} Takes care of all the steps required to generate, build, package, and deploy native C\# web applications on top of a particular stack (e.g., for a Windows Server using SQL Server this will be ASP.Net and SQL code). 
The compiled application is then deployed to the Application Server.

\textbf{Application Server.} Runs on top of IIS. The server stores and runs the developed
applications 

\textbf{Service Center.} Provides a web interface for sysadmins, managers and operations teams. Connections to external databases are configured here.


\textbf{Integration Studio.} A desktop environment, targeted at technical developers, used to integrate external libraries, services, and databases.

\subsection{Data Access}

One of the most common operations in data-driven applications is to fetch and display data from the database. 
The OutSystems Platform models the data access through the concept of Entities, elements that enable information to be persisted in the database and the implementation of a database model, following the relational model to represent data and its relationships.
A visual editor that allows development teams to query and aggregate data visually is also provided, so that developers with any skill set can work with the complex data needed for any application. 

Using this platform, developers can create integrations of local and external data sources without having to write custom code, significantly reducing time and effort, and eliminating errors. 
OutSystems integrates natively with several of the major relational database systems: SQL Server, SQL Azure, Oracle, MySQL, and DB2 iSeries. This allows the development of applications that access data in external databases using the OutSystems entity model without having to worry about data migration.
Integration with external databases involves:
(1) In the Service Center, defining a connection to the external database;
(2) In the Integration Studio, mapping tables or views in the external database to entities in an extension module;
(3) In the Service Studio, referencing and using the extension in an application.

\section{Data stores}
\label{sec:nosql}

%

NoSQL data stores, in general, forgo the requirement for a strict structure,
sacrificing query processing capabilities for flexibility and efficiency. While SQL
became an ANSI standard, allowing queries to be executed in compatible RDBMS
with few idiomatic changes, NoSQL data stores offer a variety of data models,
also supporting different data types and query capabilities. Additionally, most support heterogeneity
at each hierarchical level, allowing, for example, rows of a given table to
have different attributes. Values can be any type of object: For document
stores, values can be semi-structured JSON documents, for example; for
wide-column data stores, values can be variable sets of columns. This
flexibility harms predictability, as additional knowledge (metadata) of what is
being stored is required, as well as the ability to handle similar data items
with irregular structures.



NoSQL data stores have initially provided only minimal query capabilities. In
fact, pure key-value stores offer only minimal get, put, and delete operations.
However, query capabilities in NoSQL data stores have been enriched over time: by embedding different subsets and extensions of SQL, as is the case
with Cassandra, Couchbase, Hive and Cloudera; 
by evolving increasingly complex
native query languages out of initially simplistic query mechanisms, as is the
case with MongoDB and Elasticsearch;
and by actually offering complex declarative
languages, that do not extend SQL but are tailored to their data models, as is
the case with Neo4j.  
In all of these, the increasing complexity and expressiveness of the language
means that these have actually evolved to have query engines that, to some extent,
perform query optimization and execution.  

The selection of data stores to be analysed in the context of this work was based on two criteria. First, variety, to include the most common NoSQL models, such as document-based, wide-column, pure key-value and graph-based. Second, utility, targeting the most popular data stores, in each class, considering the client base of the OutSystems platform.

The minimum requirements for NoSQL data stores to be compatible with the OutSystems platform are:
(1) the ability to expose the database schema, to enable the platform user to select the appropriate databases, tables, attributes or equivalent constructs, explored in Section~\ref{sec:metadata}; (2) a compatible model (and syntax) for expressing queries considering filtering operators such as selection and projection, ordering and grouping operators such as sort by and group by, and aggregation operators for counting, summation and average; a compatible structure for consuming query results, currently restricted to scalar values or flat structures, i.e., no nested data.

Additionally, it would be desirable to: enable visual construction of queries based on online partial results; have the ability to take advantage of NoSQL-specific features; and provide client-side query simplification.

While the platform also supports a hierarchical model currently geared towards consuming data from REST endpoints, data manipulation using the relational model is better supported, as the current query construction interface uses a tabular representation for data. Also, it is a more natural fit for the guiding principles of a low-code platform as it is more likely to be familiar to developers. Focusing on this route for integration, for each examined data store, we evaluate: (1) the available query operations, including traversal of data, filtering, and computing different aggregations (Section~\ref{subsubsec:queryop}); (2)  how such operations can be combined (Section~\ref{subsubsec:queryconstruc}); and (3) how to map such operators and compositions to the OutSystems model (Section~\ref{subsec:discussion}).

\

\subsection{Query Operators}
\label{subsubsec:queryop}
The fitness of the relational model for interfacing with each NoSQL data store rests on: (1) the ability to express a given operation in the data store's model and query language, even if some translation is required and (2) on results being either scalar values or convertible to a tabular form.

In order to use a relational notation with both document-based and row-based models supporting nested structures (including collections) these must be flattened. One way to do this is to promote the fields of the nested structure to outermost attributes and unnest lists, creating a separate row for each element of the list.

%

\begin{figure*}
\begin{Verbatim}[fontsize=\small]
{
 _id: 		(hidden)
 id: "store::1",           
 location: "Braga",
 sells: [
  { widget: { id: "Widget1", color: "red" }, qty: 5 },
  { widget: { id: "Widget2", color: "blue" }, qty: 2 },
 ]
}
\end{Verbatim}
\caption{Example MongoDB document for a widget store.}
\label{fig:mongoeg}
\end{figure*}

\begin{figure}[t]
 \centering
 \includegraphics[width=\columnwidth]{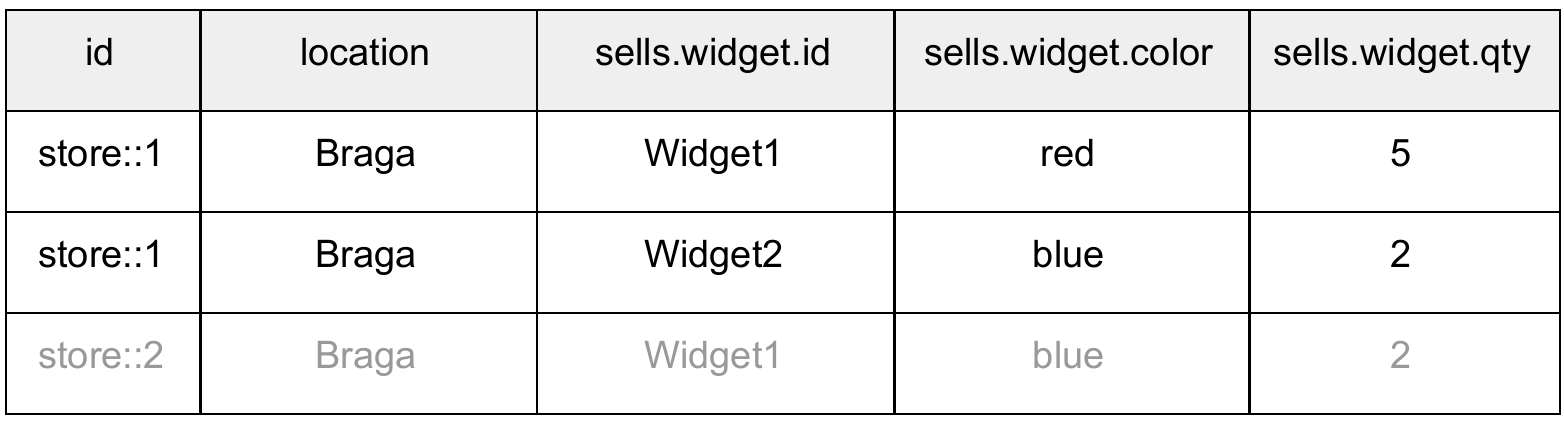}
 \caption {Representation of a MongoDB document in a relational model.}
 \label{tab:relmongo}
\end{figure}

As an example, consider the MongoDB document shown in Figure~\ref{fig:mongoeg}.
This document can be modelled as a relation, where the name of each attribute is prefixed with a path to reach it in the original model.
The result from converting the document to this model is depicted in Figure~\ref{tab:relmongo}. Data from other documents in the same collection could be added as rows. If the schema is partial (or probabilistic), some rows will likely be missing some attributes.
Naturally, this step requires the schema to be retrievable, whether explicit or inferred, a concern addressed in Section~\ref{sec:metadata}.

%

This method can be extended to graph databases, for acyclic schemas. 
Using Neo4j as an example, considering the information that can be retrieved regarding node and edge properties, a relational schema could be exposed with
a relation per node type, with properties as attributes and a relation per edge type (relationship type in Neo4j).
However, taking full advantage of the capabilities of the graph model may not be as straightforward, such as, for example, using the length of the path in a query.

Consider, instead, a key-value data model with collections, such as Redis. While extracting a high-level schema for the stored data from Redis is not generally feasible, it can be done on a per key basis, first by listing the keys and querying the type of the associated value. Each value type can be straightforwardly converted to a tabular format, except for HyperLogLogs, a probabilistic data structure used to count unique items.
%
%
Selection (filtering rows/records) and projection (filtering co\-lumns/\-at\-tributes) operators are generally supported in NoSQL data stores, even if projections require defining paths on possibly nested structures. 

\begin{figure}[p]
 \centering
 \includegraphics[width=\columnwidth]{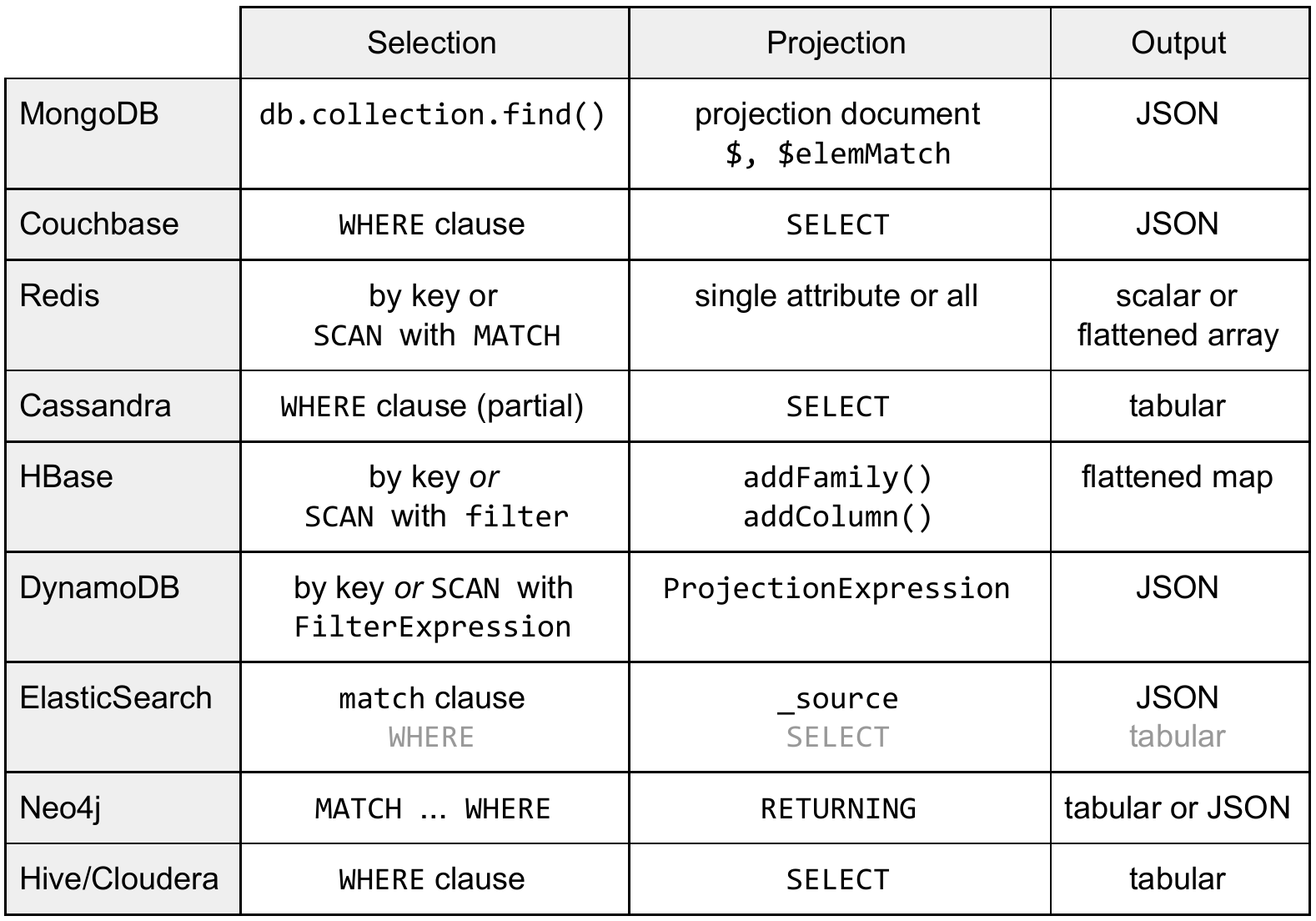}
 \caption{Selection and Projection operators in NoSQL data stores.}
 \label{fig:selandproj}
\end{figure}

The commands/keywords to make use of a given feature in the particular data store are presented, in Figure~\ref{fig:selandproj}, as a command where available, or a short explanation of what is available, where convenient.

%
MongoDB is a JSON-centric database system, using it both as its data model and as its query language. It has thus been highly successful for Web applications where used together with JavaScript. Objects are grouped in collections, that are stored in databases. Each object is identified in the context of a collection by the \texttt{\_id} field for direct access. This field can be provided by the application with any relevant content, or is otherwise automatically generated with a unique ObjectId. MongoDB can store relational data by introducing reference fields that hold the value in the\texttt{\_id} of the referenced document. Nonetheless, MongoDB encourages denormalization and storing related entities as a hierarchical structure in a single document. The native API for MongoDB is a JavaScript library, where queries are expressed as JSON documents. Similar APIs exist however for a variety of programming languages and platforms. First, this interface allows the application to insert, retrieve, and delete documents directly referenced by their \texttt{\_id}. Second, it allows searching for documents providing a document for matching. At its simplest, this includes key-value pairs that need to exist in the target documents. It allows also several relational and boolean operators to be specified with keywords prefixed by the \texttt{\$} symbol, allowing complex expressions to be composed. In update operations, the new value can also be computed by using a set of functions accessed also with \texttt{\$} keywords.
The \texttt{db.collection.find()} method retrieves all documents that match a set of criteria defined over the fields of the document. While, by default, all fields of matching documents are retrieved, the set of fields to be projected can be defined in a projection document. Projection of elements in nested arrays can also be controlled.
Couchbase is, like MongoDB, a document store that uses JSON as its data model. It provides a simple key-value interface and N1QL, a rich SQL-like query language. The outermost data organization units are buckets, which group key-value pairs (items). Keys are immutable unique identifiers, mapped to either binary or JSON values. Like MongoDB, relations between documents can be created either by referencing other documents by key or through embedding, creating a hierarchical structure. Selection and projection are similar to SQL's but dealing with nested structures such as arrays, requires explicitly unnesting these, using the \texttt{UNNEST} clause.

Redis is an in-memory key-value store using strings as keys and where values can be strings or data structures such as lists, sorted or unsorted sets, hash maps, bit arrays and HyperLogLogs, probabilistic data structures used to count unique items, trading in precision for space savings. Querying over values is not supported, only queries by key. The alternative is costly: a full scan, returning key-value pairs that match a specified condition. In terms of projection, either a single attribute is retrieved or all are. Redis is used primarily for caching, with optional persistence. Another usage pattern is to use Redis for write-intensive data items, using a backing data store for other data.

Apache Cassandra is a wide-column store inspired by BigTable. Cassandra, however, supports super column families as an additional aggregate. It supports CQL, a SQL-like query language. Queries that use relation traversal, i.e. the equivalent to JOINs in the relational model, require either custom indices or secondary indices on would-be JOIN attributes to be available. Index creation is asynchronous.

Apache HBase is also a wide-column store, modelled on Google BigTable and implemented on the Hadoop stack. HBase does not constrain data stored in keys, qualifiers (attributes), and values. Data items have therefore to be serialized to byte arrays before storage and deserialized back upon retrieval. HBase natively offers a simple Java client API and no query language. Selection is implemented by either retrieving a single or multiple rows, given the keys and an optional upper bound on the timestamp, or by scanning a range of rows, given (optional) start and stop keys, optionally providing a filter expression comparing a value to a provided constant. Projection is supported through the definition of a subset of qualifiers to be retrieved.

Amazon DynamoDB is a large scale distributed data store offered as part of Amazon Web Services.
Data is stored in tables, where rows are identified by simple or composite primary keys. If using a simple primary key, it serves as the partition key. Composite primary keys are limited to two attributes, where the first attribute is also used as a partition key and the second is used for sorting the rows. Selection and projection capabilities are similar to those described for HBase. Answering queries that use relation traversal efficiently requires creating secondary indexes. However, the number of indices that can be created is limited. The costly alternative is to use table scans and combine data client-side.

Elasticsearch is a document store focusing on text indexing and retrieval, primarily developed as a distributed version of the well known Lucene text indexing library. It uses the well known JSON format as its base data model and groups JSON objects in collections called ``indexes". Each object in an index has a unique identifier used for direct access. It is expected that each of these indexes contains a fairly homogeneous set of objects, with mostly the same attributes and being used for the same purpose. Elasticsearch allows a very limited form of relational data by introducing document parent-child relationships. This allows splitting very large documents that contain many sub-elements and would be inefficiently managed, while maintaining the ability to traverse them together in some operations. This possibility has however many restrictions and the documentation emphatically discourages its generalized use. Selection and projection are supported either in the native API or an experimental SQL-like API.

Neo4j is a graph data management system based on the Java platform. The graph data model and operations make it a very different alternative to most other well known NoSQL systems.
Information is represented as a directed graph defined as a set of named nodes and a set of edges, each connecting a source node to a target node. Additional information can then be attached to nodes and edges in one of two ways:
as a single tag, that is a symbolic identifier with particular significance as metadata and for indexing; or as a set of named properties, that can have various types and hold arbitrary data. Values of types attached to properties, that can be manipulated in queries, can be both primitive types such as numbers or strings, but also composite values such as lists and maps. This means that the data attached to nodes and edges can be nested in structures as typical of JSON documents. The main interface to Neo4j is Cypher, a declarative graph query and manipulation language. It natively supports graph concepts such as paths and matching on graph structure. Along with SQL-like selection and projection capabilities on the data attached to nodes and edges, attributes such as the length of a path can also be specified.

Apache Hive and Cloudera Impala are analytic query engines for a SQL-like query language using, preferably, data stored in HDFS files. Hive is built on top of Hadoop and translates queries to MapReduce jobs. Impala translates queries to a traditional Volcano-style parallel execution plan. They share the same client API, most of the language, and some of the infrastructure, namely, for storage and metadata. Data can be organized in rich hierarchical structures, that are however strongly typed. Nevertheless, processing is done in terms of flat rows.

Figure~\ref{fig:selandproj} also presents the output format for query results. 
While some of the analysed data stores present results in document or row-based data models with nested structures, these can be easily converted to a tabular format using the already discussed techniques, fulfilling requirement (2).

\begin{figure}[p]
 \centering
 \includegraphics[width=\columnwidth]{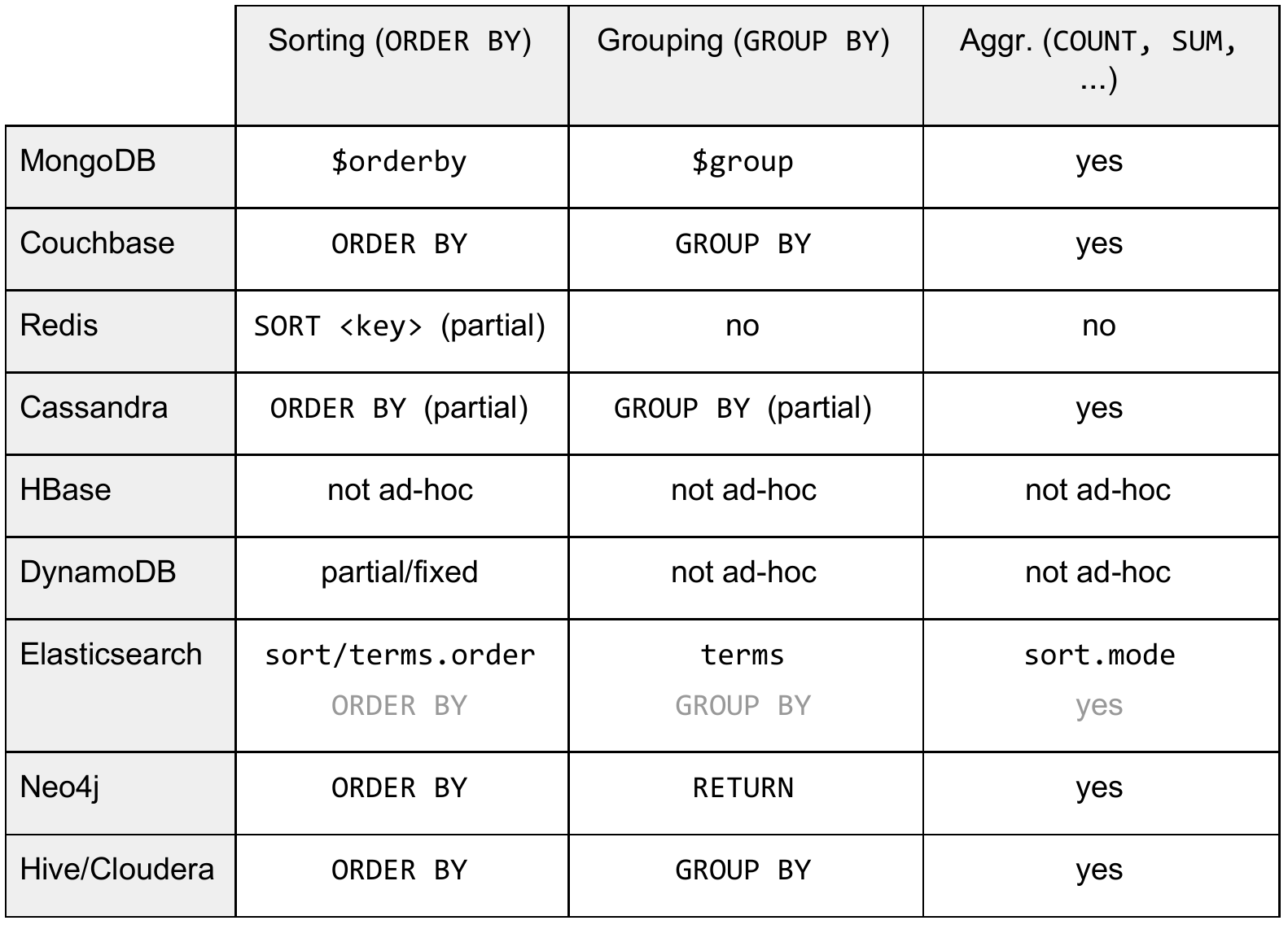}.
 \caption{Sorting, Grouping and Aggregation operators in NoSQL data stores.}
 \label{fig:sortgroupagg}
\end{figure}

Regarding sorting, grouping and aggregation operators, Figure~\ref{fig:sortgroupagg} shows the commands/keywords to make use of the given feature in the particular data store, presented as a command where available, or a short explanation of known limitations where convenient. Redis only sorts elements in a data structure (list, set, sorted set) stored for a specific key. Again, support for SQL-like commands in Elasticsearch is experimental.
Cassandra supports sorting only on one of a table's attributes, the first clustering column, while grouping can only be done using the partition key. 
DynamoDB has similar restrictions on ordering. Aggregates require creating indexes/materialized views, precluding ad-hoc queries.
Results in HBase are sorted in a fixed order, using the row key for the outermost comparison. Aggregates can be implemented in HBase using coprocessors.
Remaining data stores implement sorting, grouping and aggregation operators that are compatible with SQL's, indicated in the column header.

\subsection{Query Construction}
\label{subsubsec:queryconstruc}

Elementary operators such as projection and selection can be combined into complete query expressions. On one extreme, data stores with limited query capabilities allow only simple fixed combinations of supported operators. For instance, HBase allows projection and selection to be optionally specified for each query. On the other extreme, data stores allow arbitrary combinations of operators to be submitted and often perform an optimization step, reordering and transforming the proposed expression for efficient execution. This section focuses on the latter, namely on the MapReduce, pipeline and tree query construction paradigms.

MapReduce is a particular query construction approach as it always assumes an intermediate grouping operation. Briefly, it works as follows:
The first step, Map, allows an arbitrary function to translate each data item in an input collection to zero or more output keys and values, usually, to filter and transform data.
In a second step, data items are grouped by key and Reduced, allowing an arbitrary function to process the set of values associated with each key, usually to aggregate them.
Often, multiple map and reduce stages can be used, thus allowing more complex queries that group by different criteria in different stages. 
MapReduce jobs are mostly used to asynchronously perform table scans and build secondary indexes, that enable these stores to offer querying capabilities that are expected of the relational model: materialised views, aggregations, etc. MapReduce can be used to execute relational equi-JOINs by using the join key for grouping, as generating all possible matches with a reducer. In this perspective, consuming MapReduce results can be done within the relational model, as long as these come in (or are easily translatable to) tabular form, possibly with an inferred schema.
First, MapReduce focuses on analytical workloads, that read most of or all input data, resulting in poor interactive performance. Moreover, limiting the specification of MapReduce tasks to SQL may be undesirable.  An argument can be made for considering that the specification of MapReduce jobs requires sufficient expertise to fall outside of the typical utilization of the platform, making it available through a low-level extension. 

A more general query construction approach is to allow arbitrary operators to be chained in a linear pipeline. The first operator reads data, often with an implicit selection and projection. Each operator does some transformation, such as grouping and aggregation, and sends data to the next.
An example of this approach is provided by MongoDB's aggregation pipeline. Figure~\ref{fig:mongoquery} demonstrates how this approach can be used over a denormalized schema, in which all data for a store is kept in a single document:

\begin{figure*}
\begin{Verbatim}[fontsize=\small]
{ aggregate: "stores", pipeline: [
        { $match: { location: { $eq: "Braga" } } },
        { $project: { id: 1, location: 1, sells: 1 } },
        { $project: { expr000: "$sells", expr001: "$id" } },
        { $unwind: "$expr000" },
        { $project: { sid: "$expr001", ITEM: "$expr000.widget.color" } },
        { $match: { ITEM: { $eq: "red" } } },
        { $project: { sid: 1 } }
    ], cursor: { batchSize: 4095
    }, allowDiskUse: true, $readPreference: { mode: "secondaryPreferred"
    }, $db: "storesdb"
}
\end{Verbatim}
\caption{Pipeline query construction.}
\label{fig:mongoquery}
\end{figure*}

While this is not the most natural way to express this query in MongoDB, it is however a simple example of an aggregation pipeline. 
The query first filters stores located in ``Braga" and projects only the required fields. Then it unwinds the ``sells" array, producing a document for each item sold. It then projects the required fields and filters widgets with color ``red".
This approach has several advantages. First, it can express a large subset of all queries possible with SQL. In fact, the linear chain of operators precludes only JOIN operations where more than one data source would be required, and some sub-queries, i.e., mainly those that are equivalent to JOIN operations.
Second, the pipeline transformation of data is intuitive, as the same paradigm is used in various forms in computer interfaces. 
Finally, data stores supporting this paradigm also tend to include query optimizers, that reorder operators to produce an efficient execution plan. This means that a query can be incrementally built without being overly concerned with its efficiency.

The most general approach is to allow arbitrary data query expressions that translate to operator trees. This allows expressing relational JOIN operations between data resulting from any two sub-queries. It also allows a general use of sub-queries whose results are used as inputs to other operators.
The main advantage of this approach is that it is what has traditionally been done with SQL databases. Ironically, NoSQL data stores therefore increasingly provide SQL-like dialects and query optimizers. This allows existing low-code query builders designed for SQL databases to be reused with minor modifications, mainly, to what functions and operators are available.

\begin{figure}[p]
 \centering
 \includegraphics[width=\columnwidth]{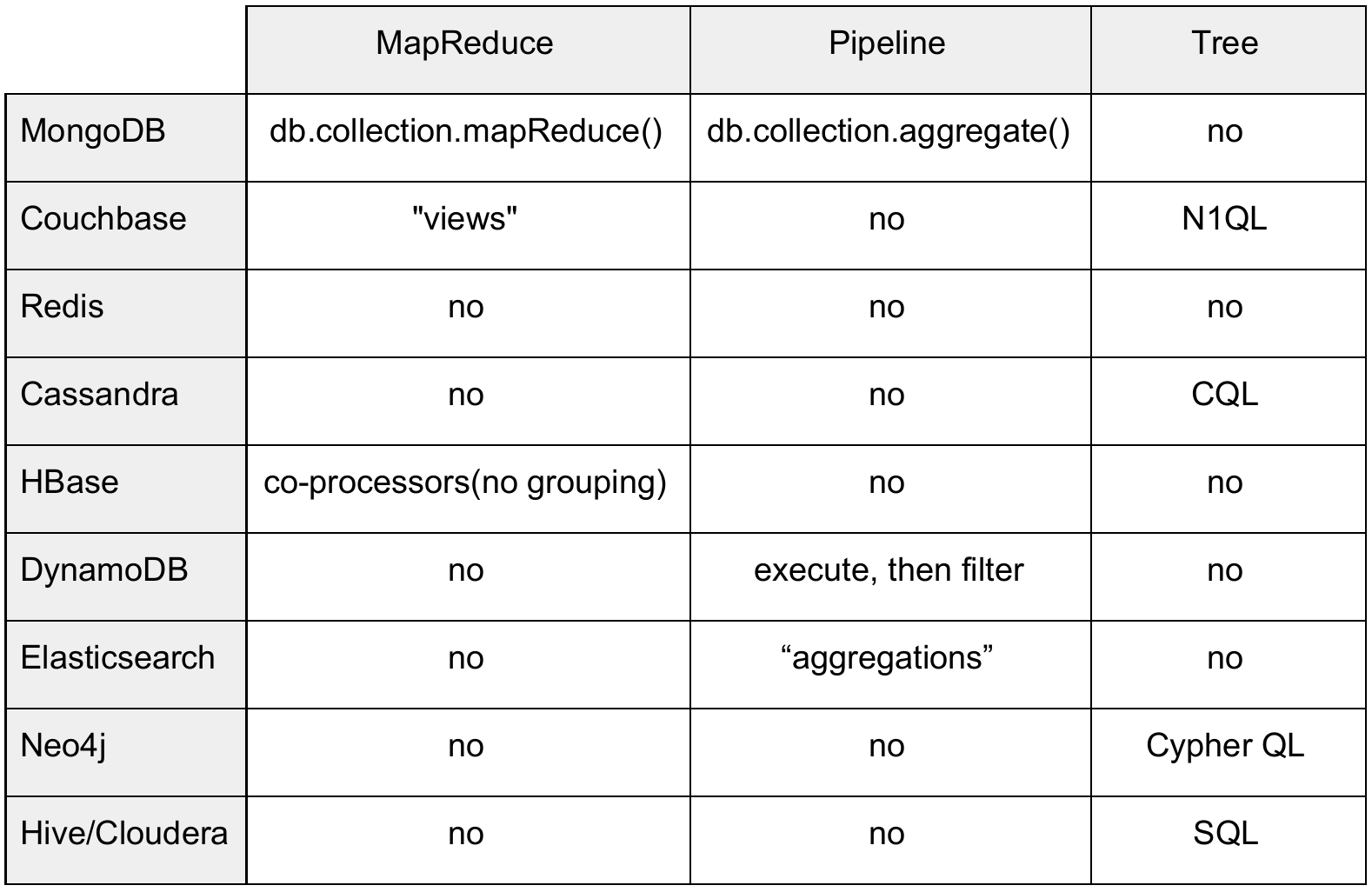}.
 \caption{Query construction paradigms in NoSQL data stores.}
 \label{fig:querycons}
\end{figure}

Figure ~\ref{fig:querycons} summarizes the supported query construction paradigms per data store, presenting either the method or concept that provides the capability.

Redis supports only a key-value interface, therefore, having no query construction capabilities.
Amazon's DynamoDB supports only simple queries by primary key (or a range thereof) with optional filtering applied just before the data is returned. 

\subsection{Discussion}
\label{subsec:discussion}

The main conclusion of the analysis is that the relational model is to a large extent able to map
enough of NoSQL data stores to fit the OutSystems' low code platform.  This means that the current aggregation model, designed primarily for relational data, can be adequately used to describe most query operators and construction schemas found in NoSQL data stores. This is a consequence of two aspects.
First, of the completeness of the relational model, that can map most operators and query construction approaches found in NoSQL data stores, requiring only an extension to support nested data. Second, of the trend for NoSQL data stores to increasingly provide SQL support, in particular, by using unwind/unnest operations to deal with denormalized and nested data.

An exception to this is the ability to directly describe a MapReduce operation (MongoDB, Couchbase, HBase). The MapReduce paradigm fits in this vision as a background workhorse to pre-process data, e.g. creating secondary indexes or materialized views, to endow NoSQL data stores with sorting, grouping or aggregation operations required for approximating SQL support. Some data stores provide this functionality automatically, which others require user intervention. In any case, as long as the results of a MapReduce job can be converted to a nested tabular format, these can be consumed in the context of the relational model.
When not supported by the data store, some SQL features can be implemented with client-side processing or a middleware solution. For example, projection can be fully supported in Redis with client-side filtering before returning data to the platform user. 
A second exception is needed to take advantage of specific NoSQL features. For example, to take advantage of features such as the added querying features the graph model brings Neo4j, more expertise than intended for using a low-code platform may be required.  Still, this can be supported by enabling advanced users to edit the query itself. 
On a different note, Redis exposes its data as simple data structures. It might make sense to allow the user to compose several of these structures into a materialized view, providing, in fact, an external schema for the stored data. 

To summarize the expected development effort:
\begin{enumerate}
\item There is no need to modify (any of these) NoSQL data stores for integration with the OutSystems platform.
\item Client-side or in-middleware computation should be limited to: (a) when needed, converting results to a tabular format (b) when needed, filling in for operators that are either missing from the data store or cannot be used ad-hoc; and (c) convert SQL to native query languages or SQL-like dialects.
\item Add support for nesting and unnesting operations to the OutSystems platform (see Figure~\ref{fig:selandproj}).
\end{enumerate}

\section{Schema Discovery}
\label{sec:metadata}
Offering a uniform view of unstructured data requires data types to be well defined and the definition of a schema to which the data must conform.
A schema is a formal description that exposes the underlying structure of data.

Currently, for integration with a data source, the platform must be able to list available databases, list tables per database and to retrieve a schema for the data.
Providing a schema to the platform is key to the low-code approach for data integration, as presenting the schema to the user, along with data samples, is required for queries to be defined visually.

For relational databases, the schema consists of relations or entities (typically represented as tables), their attributes (including data types) and, potentially, the cardinality associated to the referential relationships between them. For NoSQL databases based on a document model, using MongoDB as an example, the schema for a database should include existing collections, document fields for each collection, including data types and, potentially, referential relationships. However, while relational schemas are flat, strict and deterministic, the flexibility of adding documents with different fields to the same collection, having fields with the same name holding values of different types and the ability to define nested structures, make it harder to elicit a complete and accurate schema of the data. 

\subsection{Third-party tools}

\begin{figure}[p]
 \centering
 \includegraphics[width=\columnwidth]{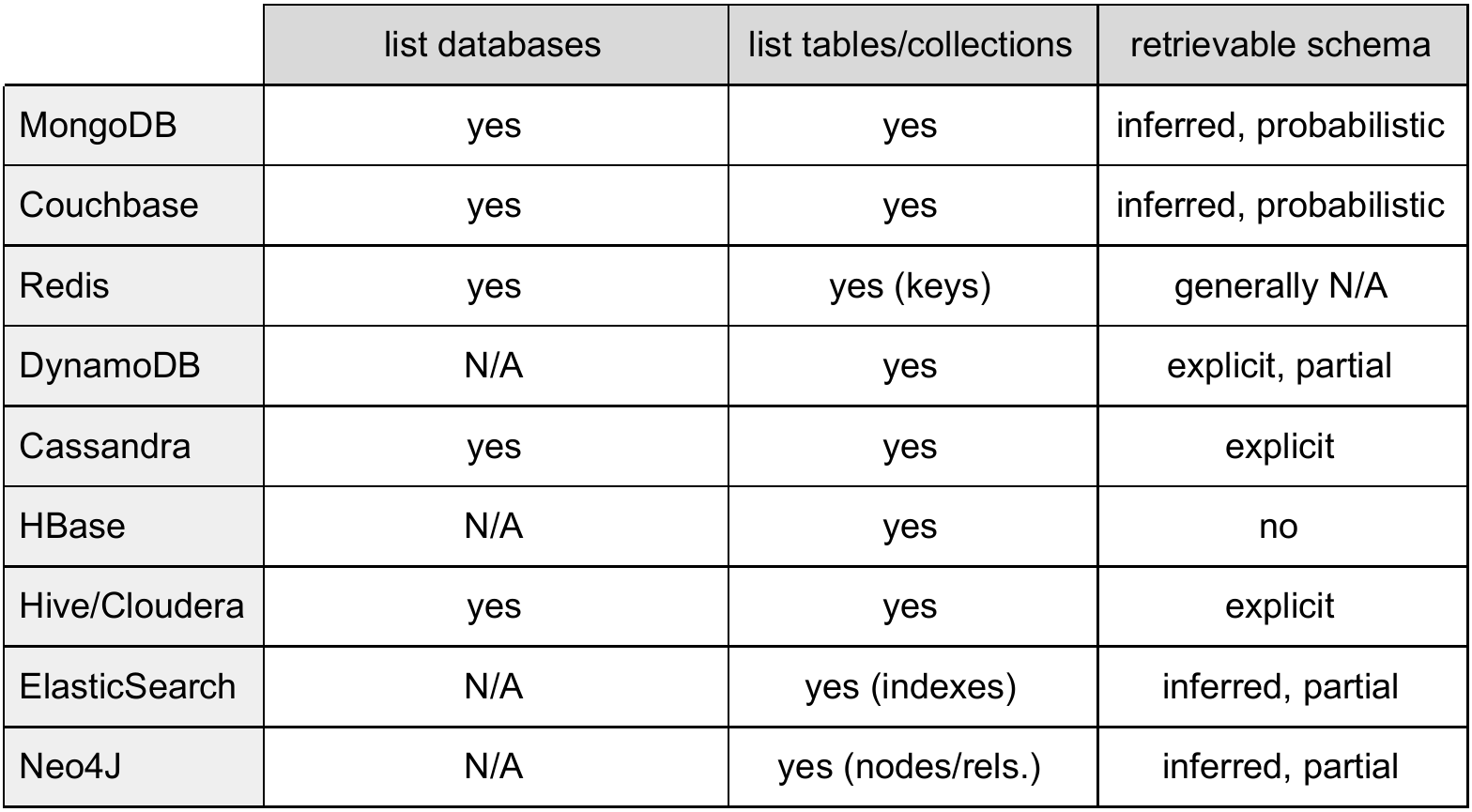}.
 \caption{Retrievable schema-related metadata from each NoSQL data store.}
 \label{fig:schema}
\end{figure}

Figure~\ref{fig:schema} summarizes the schema-related metadata that can be retrieved or inferred from each of the analysed data stores, using currently available tools. Cells marked N/A indicate there is no comparable concept in the store's data model. Partial schema inference refers to cases where nested structures are opaque.

A wide range of NoSQL data stores explicitly store and provide metadata, a confirmation of the JSON data model, that allows nested data collections, as a common trend for data structuring and representation. There are however minor differences in what data types can be attached to values and even how much nesting is allowed.

Second, NoSQL data stores vary widely in which metadata can be obtained. On one end, systems such as Cassandra, Hive, and Cloudera Impala store and enforce a user-defined schema that unambiguously describes data.
On the other end, systems such as HBase do not provide any support for setting or enforcing a schema. As they store just arbitrary byte sequences, the schema cannot also be easily inferred. They are however frequently used with third-party query mechanisms, that provide the missing metadata. HBase, for example, is often used in the Hadoop stack with Hive.
In between, systems such as MongoDB, Couchbase, Elasticsearch and Neo4j allow the schema to be inferred from currently stored data and even provide optional mechanisms to enforce it. For example, using MongoDB it is possible to define a set of rules over the attributes of documents belonging to a given collection. Validation is defined at the collection level and conditions can require a document to contain a given set of attributes, with a given type, set attribute-level boundary conditions, or require values to match a given regular expression. In effect, introducing validation limits the variability of the structure of JSON documents.

Mechanisms used to infer schema from MongoDB and Couchbase can, in principle, be applied to other schema-less data stores, by adapting existing tools or by implementing similar ones. In short, tools such as \texttt{mongodb-schema}\footnote{\url{https://github.com/mongodb-js/mongodb-schema}}analyze a sample of documents stored in a given collection and provide, as outcome, a probabilistic schema. Fields are annotated with a probability according to how frequently these occur in that collection's sampled documents. Fields are also annotated with a set of data types: each data type is itself annotated with a probability according the mapping's occurrence in the sampled documents. There is also \texttt{mongodrdl}\footnote{\url{https://docs.mongodb.com/bi-connector/current/reference/mongodrdl/}}, a tool for inferring a relational schema from a MongoDB database or collection, which, however, in our experiments, fell short of accurately representing some relationships between unnested array elements and top-most attributes.

A similar concern holds for using data sources with polyglot query engines, as these enable expressing data processing operations over systems that expose multiple native data models and query languages.

Dremio OSS performs schema inference, but treats nested structures as opaque and, therefore, does not completely support low-code construction of unnesting operations, in the sense that the user still needs to explicitly handle these. Still, it provides the ability to impose a table schema ad-hoc or flexibly adapt data types which is a desirable feature for overriding incorrect schema inference. 

With PostgreSQL FDW, it is possible to declare tables for which query and manipulation operations are delegated on adapters. The wrapper interface includes the ability to either impose or import a schema for the foreign tables. Imposing a schema requires the user to declare data types and structure and it is up to the wrapper to make it fit by using automatic type conversions as possible. If this automatic process is not successful the user will need to change the specified data type to provide a closer type match. The wrapper can also (optionally) advertise the possibility of importing a schema. In this case, the user simply instructs PostgreSQL to import meta-data from the wrapper and use it for further operations. This capability is provided by the wrapper and currently, this is only supported for SQL databases, for which the schema can be easily queried.
Furthermore, PostgreSQL FDW can export the schema of the created foreign tables.
Both for Dremio and PostgreSQL, limitations in schema imposition/inference do not impact querying capabilities, only the required talent to use the system.  For PostgreSQL FDW, this can be mitigated by extending adapters to improve support for nested data structures, integrating schema inference/extraction techniques, as proposed and described in Section~\ref{subsec:implementation}.

\subsection{Improving current tools}
Here, we briefly consider research results, for which runnable systems are not generally available, but which nonetheless can contribute relevant ideas and techniques.
Generic schema inference/discovery focuses on discovering, for a set of semi-structured data, how these can be generically represented as sets of attributes (entities) of a given type (or set thereof), optionally identifying relationships between entities (e.g., references) and constraints (e.g. that a given attribute is required to be non-null).
Proposals are typically motivated by the necessity of designing client applications that can take advantage of a semi-structured data source for which a schema is unknown with earlier work focused mainly on XML data.
One approach, is to involve the user in schema discovery by exposing generated relational views of the JSON data to users so that these can help clusters records as collections and mark or validate relationships between entities \cite{spoth2018schemadrill}. While the goal of this particular work is to ultimately provide a flat relational schema of the JSON data, concerns such as minimising the number of collections in the final schema and introducing relationships between entities might have a significant impact in the effectiveness of querying JSON data, an aspect that is not assessed by the authors. This type of approach seems to be better suited for data exploration than for integrating data sources with applications in a low code setting.

It has also been proposed that machine learning can be used to generate a relational schemas from semi-structured data \cite{discala2016automatic}. However, while the authors did perform a preliminary assessment of the query performance on the generated schemas, queries were performed on the data loaded onto a relational database. Results from this assessment do not necessarily hold when queries (operations) are pushed down to a NoSQL store. 

A commonly identified pattern is that the same conceptual object can be represented by documents that differ in a subset of fields, or have fields with different types, as a consequence of the schema-less nature of semi-structured data. This effect  can be captured as coalescing these slightly different effective schemas as different versions of the same document schema. In \cite{ruiz2015inferring}, the authors propose a method based on model-driven engineering to do just that.

A significantly different approach for schema discovery is to analyse application source code to discover the schema implicitly imposed by the application on schema-less data sources. In \cite{castrejon2013exschema}, the authors propose such a method for applications that use relational and NoSQL data sources. While currently out-of-scope, it might be interesting to offer this type of capability to ease the migration of applications to the OutSystems platform.

\section{Architecture}
\label{sec:prop}

Considering the conclusions from surveying a variety of NoSQL systems, in particular, regarding their supported data and query models, we describe the architecture for a polyglot data access layer for a low-code application platform, and then discuss a proof-of-concept implementation based on existing open source components.


Our proposal is based on two main criteria. First, how it contributes to the vision of NoSQL data integration in the low-code platform outlined in Section~\ref{sec:intro} and how it fits the low-code approach in general. Second, the talent and effort needed for developing such integrations and then, later, for each additional NoSQL system that needs to be supported.

We can consider two extreme views. On the one hand, we can enrich the abstractions that are exposed to the developer to encompass the data and query processing models. This includes: data types and structures, such as nested tuples, arrays, and maps; query operations, ranging from general purpose data manipulation (e.g., flattening a nested structure) to domain-specific operations (e.g., regarding search terms in a text index); and finally, where applicable, query composition (e.g., with MapReduce or a pipeline).

This approach has however several drawbacks. First, it pollutes the low-code platform with a variety of abstractions that have to be learned by the developers to fully use it. Moreover, these abstractions change with support for additional NoSQL systems and are not universally applicable. In fact, support for different NoSQL systems would be very different, making it difficult to use the same know-how to develop applications on them all. Finally, building and maintaining the platform itself would require a lot of talent and effort in the long term, as support for additional systems could not be neatly separated in plugins with simple, abstract interfaces.

On the other hand, we can map all data in different NoSQL systems to a relational schema with standard types and allow queries to be expressed in SQL. This results in a mediator/wrapper architecture that allows the same queries to be executed over all data regardless of its source, even if by the query engine at the mediator layers.

This approach also has drawbacks. First, mapping NoSQL data models to a relational schema requires developer intervention to extract the view that is adequate to the queries that are foreseen. This will most likely require NoSQL-specific talent to write target queries and conversion scripts. Moreover, query capabilities in NoSQL systems will remain largely unused, as only simple filters and projections are pushed down, meaning the bulk of data processing would need to be performed client-side.

Our proposal is a compromise between these two extreme approaches, that can be summed up as: support for nested data and its manipulation in the abstractions shown to the low-code developer, along with the ability to push aggregation operations down to NoSQL stores from a mediator query engine, will account for the vast majority of use cases. In addition, the ability to embed native query fragments in queries will allow fully using the NoSQL store when talent is available, without disrupting the overall integration. The result is a polyglot query engine, where SQL statements are combined with multiple foreign languages for different NoSQL systems.

\begin{figure}[t]
\centering
\includegraphics[width=1\columnwidth]{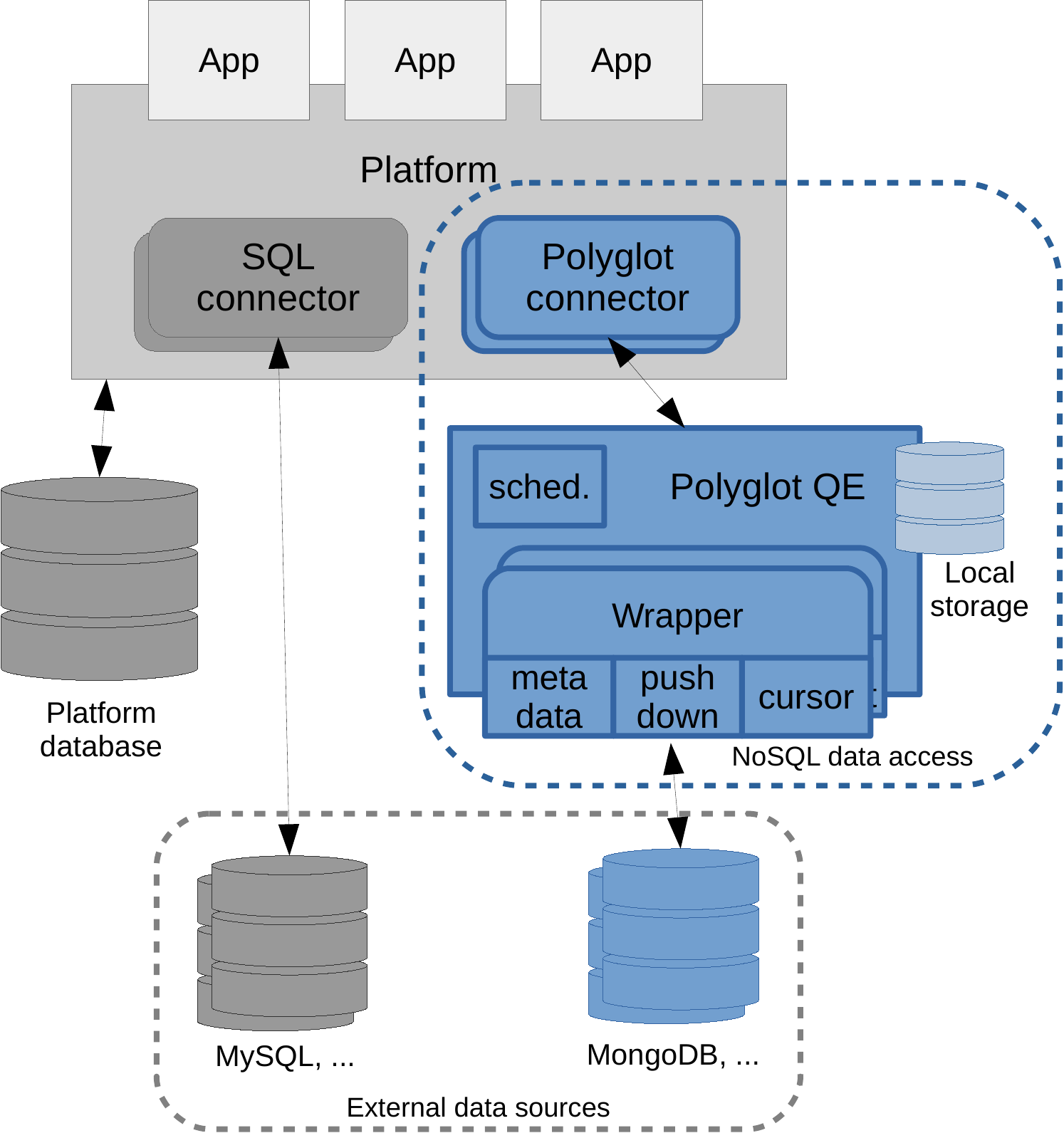}
\caption{Architecture overview}
\label{fig:arch}
\end{figure}

The proposed architecture is summarized in Figure~\ref{fig:arch}, highlighting the proposed NoSQL data access layer. To the existing OutSystems platform, encompassing development tools and runtime components, we add a new \emph{Polyglot connector}, using the Database Integration API to connect to the \emph{Polyglot Query Engine (QE)} through standard platform APIs.
The Polyglot QE acts as a mediator. It exposes an extended relational database schema for connected NoSQL stores and is able to handle SQL and polyglot queries.

For each NoSQL Store, there is a \emph{Wrapper}, composed of three sub-components:
 \emph{metadata} extraction, responsible for determining the structure of data in the corresponding store using an appropriate method and mapping it to the extended SQL data model of the Polyglot QE; a query \emph{push-down} component, able to translate a subset of SQL query expressions, to relay native query fragments, or produce a combination of both in a store-specific way; and finally, the \emph{cursor}, able to iterate on result data and to translate and convert it as required to fit the common extended SQL data model.

The Polyglot QE makes use of \emph{Local storage} for the configuration of NoSQL store adapters and for holding materialized views of data to improve response times. The \emph{Job Scheduler} enables periodically refreshing materialized views by re-executing their corresponding queries.

\subsection{Implementation}
\label{subsec:implementation}

We base our proof-of-concept implementation on open source components. In this section we start by describing how we selected those components and then describe the additional development needed to make it fit the proposed architecture.
We base our proof-of-concept implementation on open source components.
\subsubsection{Component selection}
The main component to select is the SQL query engine used as the mediator. Besides its features as a query engine, we focus on: the availability of wrappers for different NoSQL systems and the talent needed to implement additional features; the compatibility of the open source license with commercial distribution; the maturity of the code-base and supporting open source community; and finally, on its compatibility with the OutSystems low-code platform. We consider two options.

\textbf{PostgreSQL with FDW\cite{fdw}.} It is an option as it supports foreign data wrappers according to the SQL/MED standard (ISO/IEC 9075-9:2008). The main attractive for PostgreSQL is that it is a very mature open source product, with a business friendly license, a long history of deployment in production, and an unparalleled developer and user community. There is also support for .NET and Java client application platforms.
In terms of features, PostgreSQL provides a robust optimizer and an efficient query engine, that has recently added parallel execution, with excellent support for SQL standards and multiple useful extensions. It supports nested data structures both with the \texttt{json}/\texttt{jsonb} data types, as well as by natively supporting arrays and composite types. It has extensive support for traversing and unnesting them.
Regarding support for foreign data sources, besides simple filters and projections, the PostgreSQL Foreign Data Wrapper (FDW) interface can interact with the optimizer to push down joins and post-join operations such as aggregations. With PostgreSQL FDW, it is possible to declare tables for which query and manipulation operations are delegated on adapters. The wrapper interface includes the ability to either impose or import a schema for the foreign tables. Imposing a schema requires the user to declare data types and structure and it is up to the wrapper to make it fit by using automatic type conversions as possible. If this automatic process is not successful the user will need to change the specified data type to provide a closer type match. The wrapper can also (optionally) advertise the possibility of importing a schema. In this case, the user simply instructs PostgreSQL to import meta-data from the wrapper and use it for further operations. This capability is provided by the wrapper and currently, this is only supported for SQL databases, for which the schema can be easily queried.
Furthermore, PostgreSQL FDW can export the schema of the created foreign tables.
In addition to already existing wrappers for many NoSQL data sources, with variable features and maturity, the Multicorn\footnote{https://github.com/Kozea/Multicorn} framework allows exposing the Python scripting language to the developer, to complement SQL and express NoSQL data manipulation operations. 

In terms of our goals, PostgreSQL falls short on automatically using existing materialized views in queries. The common workaround is to design queries based on views and later decide whether to materialize them, which is usable in our scenario. Another issue is that schema inference is currently offered for relational data sources only. The workaround is for the developer to explicitly provide the foreign table definition.

\textbf{Calcite\cite{calcite} (in Dremio OSS\cite{dremio}).} The Calcite SQL compiler, featuring an extensible optimizer, is used in a variety of modern data processing systems. We focus on Dremio OSS as its feature list most closely matches our goal. Calcite is designed from scratch for data integration and focuses on the ability to use the optimizer itself to translate parts of the query plan to different back end languages and APIs.
It also supports nested data types and corresponding operators. Dremio OSS performs schema inference, but treats nested structures as opaque and, therefore, does not completely support low-code construction of unnesting operations, in the sense that the user still needs to explicitly handle these. Still, it provides the ability to impose a table schema ad-hoc or flexibly adapt data types which is a desirable feature for overriding incorrect schema inference. 
Also, Dremio OSS adds a distributed parallel execution engine, based on the Arrow columnar format, and a convenient way to manage materialized views (a.k.a., ``reflections''), that are automatically used in queries. Unfortunately, one cannot define or use indexes on theses views, which reduces their usefulness in our target application scenarios.

Although Calcite has a growing user and developer community, its maturity is still far behind PostgreSQL. The variety of adapters for different NoSQL systems is also lagging behind PostgreSQL FDW, although some are highly developed. For instance, the MongoDB adapter in Dremio OSS is able to extensively translate SQL queries to MongoDB's aggregation pipeline syntax, thus being able to push down much of the computation and reduce data transfer. The talent and effort needed for exploiting this in additional data wrappers is, however, substantial.
Both for Dremio and PostgreSQL, limitations in schema imposition/inference do not impact querying capabilities, only the required talent to use the system.  For PostgreSQL FDW, this can be mitigated by extending adapters to improve support for nested data structures, integrating schema inference/extraction techniques.
Finally, the main drawback of this option is that, as we observed in preliminary tests, resource usage and response time for simple queries is much higher than for PostgreSQL.

\textbf{Choosing PostgreSQL with FDW.} In the end, we found that our focus on interactive operational applications and the maturity of the PostgreSQL option, outweigh, for now, the potential advantages from Calcite's extensibility.

\textbf{Additional development} Completing a proof-of-concept implementation based on PostgreSQL as a mediator requires additional development in the low-code platform itself, an external database connector, and in the wrappers.
As examples, we describe support for two NoSQL systems. The first is Cassandra, a distributed key-value store that has evolved to include a typed schema and secondary indexes.  It has, however, only minimal ad-hoc query processing capabilities, restricted to filtering and projection. The second is MongoDB, a schema-less document store that has evolved to support complex query processing with either MapReduce or the aggregation pipeline. Both are also widely used in a variety of applications.

\textbf{Schema conversion.}
In order to support relational schema introspection, we reuse \texttt{mongodb-schema}\footnote{\url{https://github.com/mongodb-js/mongodb-schema}}, extending it to provide a probabilistic schema, with fields and types, for each collection in a MongoDB database. Top-level document fields are mapped as table attributes. When based on probabilistic schemas, all discovered attributes are included, leaving it to the user/developer to decide which attributes to consider. Nested documents' fields are mapped as top-level attributes, named as the field prefixed with its original path. Nested arrays are handled by creating a new table and promoting fields of inner documents to top-level attributes. Documents from a given collection become a line of the corresponding table (or tables).
An alternative would be to create a denormalized table, as shown in Table ~\ref{tab:relmongo}.
Notice that this is equivalent to the result of a natural join between the corresponding separate tables. However, separate tables fit better what would be expected from a relational database and thus improve the low-code experience.
It should be pointed out that viewing the original collection as a set of separate relational tables has no impact on the performance of a query with a join between these tables. The required unnesting directives, using the \texttt{\$unwind} pipeline aggregation operator are also generated and added to the table definition. We also provide the option, on by default, of adding a column referencing the \texttt{\_id} of the outermost table to all inner tables on schema generation, that can serve as an elementary foreign key.

\textbf{MongoDB wrapper.} There are multiple FDW implementations for MongoDB. We selected one based on Multicorn,\footnote{\url{https://github.com/asya999/yam_fdw}} for ease of prototyping, and change it extensively to include schema introspection and, taking advantage of aggregation pipeline query syntax, to allow push-down to work with user supplied queries.
This is greatly eased by MongoDB's syntax for the aggregation pipeline being easily manipulated by programs, by adding additional stages. 

\textbf{Cassandra wrapper.} We also use a wrapper based on Multicorn.\footnote{\url{https://github.com/rankactive/cassandra-fdw}}. In this case, we add the ability to use arbitrary Python expressions to compute row keys from arbitrary attributes, as in earlier versions of Cassandra it was usual to manually concatenate several columns. Even if this is no longer necessary in recent versions of Cassandra, it is still common practice in other NoSQL systems such as HBase. The currently preferred interface to Cassandra, CQL, is not the best fit for being manipulated by programs, although, being so simple, it can be done with relatively small amount of text parsing.

\textbf{Connectors.} We implemented custom connectors for each NoSQL store based on the original PostgreSQL connector. This allows the developer to directly pick the target data store from the platform's visual development environment \cite{outsystems:service} drop-down menu and provide system specific connection options. It also allows system specific projection and aggregation operators to be handled.

\textbf{Developer platform.} The changes needed in the platform to fully accommodate the integration are the ability to express nesting and unnesting operators in the data manipulation UI, and to generate SQL queries that contain them when using the NoSQL integration connectors. It is, however, possible to workaround this by configuring multiple flattened views of data, as needed, when the schema is introspected and imported.

\section{Use case}
\label{sec:eval}

The proposed architecture and proof-of-concept integration of the OutSystems low-code platform with NoSQL data stores is evaluated by applying it to a simple application using multiple NoSQL stores.

\subsection{Application and workload}

We use a subset of the TPC-C\,\cite{tpc:tpcc} database schema and workload as our application scenario. Note that we don't aim at evaluating database system performance and don't expect that the results obtained here are valid as such. The reasons for using TPC-C are the following: First, the database schema and operations are very well known by the database research community and in industry, making it easier to follow and understand the examples used. Second, and most importantly, we make use of \emph{py-tpcc},\footnote{\url{https://github.com/apavlo/py-tpcc}} an implementation of the TPC-C database schema and workload for multiple NoSQL (and some SQL) systems that exploits the native idioms for each of them. This provides us with a Rosetta stone that we can use to compare the results obtained with our proposal. In particular, \emph{py-tpcc} denormalizes data as fit for NoSQL systems.

Briefly, TPC-C simulates a generic wholesale supplier business, with multiple geographically distributed warehouses and associated sales districts. All warehouses stock the same set of items and each district as a corresponding set of customers. We consider only two out of the five transactions: 
 \begin{description}
   \item[Stock-Level] Determines the number of recently sold items (i.e., the latest 20 orders) with stock level below a user provided threshold.
   \item[Order-Status] Queries the delivery status for each of the items in a customer's last order, where the customer is specified either by its primary key identifier or by its name.
 \end{description}
Therefore, from the nine tables in the complete TPC-C database schema, we use only CUSTOMER, ORDERS, ORDER-LINE, DISTRICT, and STOCK. We do however create, populate, and map the entire schema.

\subsection{Experimental setup}

The implementation of TPC-C on Cassandra in \emph{py-tpcc} closely follows the relational database schema in the original specification. In particular, it uses a separate column family with the corresponding columns for each TPC-C table. However, it has two complications: First,
all data is stored as UTF-8 strings instead of using the actual data types. Second, 
the key for each column family is manually assembled by the application code by concatenating the relevant column values and padding them with zeros.
These complications make it harder to map it to a relational schema. However, these are current practice given the limitations of older versions of Cassandra (i.e., before 3.0) and thus make an excellent use case. The same practices are widely used in other key-value stores, where different application-specific methods are used to encode keys.

The implementation of TPC-C on MongoDB in \emph{py-tpcc} is a good use-case as it does not directly map each TPC-C table to a separate document collection. Instead, it maps the ORDERS for a given client as an array nested in each CUSTOMERS document, and ORDER-LINE as another array nested in each of the ORDERS.
HISTORY is also nested in CUSTOMERS, although we don't ever read or update it in Stock-Level and Order-Status transactions. Other tables are directly mapped to separate document collections, of which we use DISTRICT and STOCK. Moreover, \emph{py-tpcc} also stores all data as UTF-8 strings instead of making use of various data types supported by MongoDB.

Our experimental setup is completed with a new implementation of \emph{py-tpcc} for PostgreSQL, derived from the SQLite implementation with minimal modifications for connection setup and parameter passing. We use it to exercise the mapping of Cassandra and MongoDB through a wrapper.

\subsection{Importing the schema}

We start by importing the schema for both Cassandra and MongoDB using the \verb|IMPORT FOREIGN SCHEMA| statement. This provides a basic mapping that can now be refined.

For Cassandra, the main issue is the use of a composite key that is created by the application by concatenating strings. As Figure~\ref{fig:cassddl} shows, this can be solved by using an embedded Python snippet that is copied directly from the original application and that generates the key for direct lookup when both columns are available. A secondary issue is that converting between integers and strings at the query engine level avoids that filters are pushed down, as a reverse conversion would need to be assumed. This is solved by changing the type of columns to small integers that are then transparently converted to strings within the wrapper.

For MongoDB, the schema importer will generate, on request, the table definition in Figure \ref{fig:mdbddl} for table ORDERS, nested in documents in the CUSTOMER collection of the tpcc database, producing a usable mapping.
\texttt{mname} options reference the document fields in the backing MongoDB collection, CUSTOMER, referenced in the table options. Columns whose \texttt{mname} does not reference an ORDERS field have been added by the developer using ALTER TABLE. This method can be used if, for example, some columns don't have the expected names (e.g., \texttt{c\_id} instead of \texttt{o\_c\_id} or to add foreign keys. However, as we intend to use existing application code that assumes the SQL schema for TPC-C, we can override the definition with different names as shown in Figure~\ref{fig:mdbddl}. 
 Adding such mappings does not require any processing as the backing data is not changed. 

\begin{figure*}[t]
\begin{Verbatim}[fontsize=\small]
ALTER FOREIGN TABLE cass.district
  ALTER COLUMN key OPTIONS (composite
   'd_id,d_w_id:str(d_id).zfill(5)+str(d_w_id).zfill(5)'),
  ALTER COLUMN d_id TYPE SMALLINT,
  ALTER COLUMN d_w_id TYPE SMALLINT;
\end{Verbatim}
\caption{Coping with type conversions and a composite column in Cassandra.}
\label{fig:cassddl}
\end{figure*}

\begin{figure*}[t]
\begin{Verbatim}[fontsize=\small]
DROP FOREIGN TABLE ymdb.orders;
CREATE FOREIGN TABLE ymdb.orders (
 o_all_local numeric OPTIONS (mname 'ORDERS.O_ALL_LOCAL'),
 o_carrier_id numeric OPTIONS (mname 'ORDERS.O_CARRIER_ID'),
 o_entry_d timestamp without time zone OPTIONS (mname
'ORDERS.O_ENTRY_D'),
 o_id numeric OPTIONS (mname 'ORDERS.O_ID'),
 o_ol_cnt numeric OPTIONS (mname 'ORDERS.O_OL_CNT'),
 o_c_id numeric OPTIONS (mname 'C_ID'),
 o_d_id numeric OPTIONS (mname 'C_D_ID'),
 o_w_id numeric OPTIONS (mname 'C_W_ID')
)
SERVER ymdbserver
OPTIONS (collection 'CUSTOMER', db 'tpcc', host 'mongo',
 auth_db 'admin', user '...', password '...',
 pipe '[{"$unwind": "$ORDERS"}]'
);
\end{Verbatim}
\caption{Unnesting and renaming columns from MongoDB.}
\label{fig:mdbddl}
\end{figure*}
\subsection{Query execution}

\begin{figure*}[t]
\centering
\includegraphics[width=\textwidth]{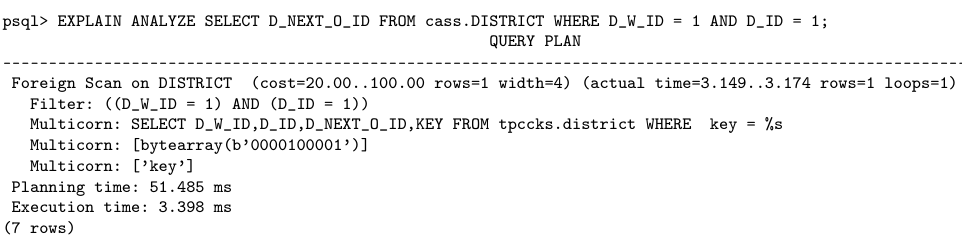}
\caption{Example of filter push-down to Cassandra with a composite key.}
\label{fig:cassexpl}
\end{figure*}

\begin{figure*}[t]
\centering
\includegraphics[width=\textwidth]{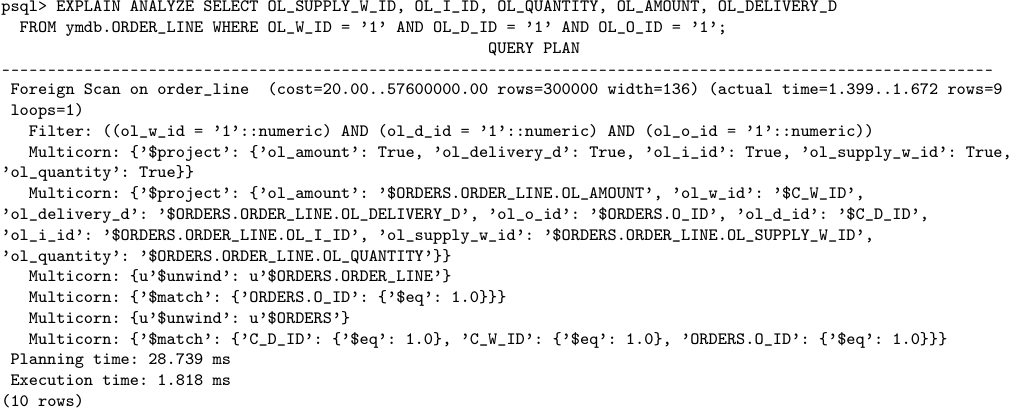}
\caption{Example of filter push-down to MongoDB on a nested table.}
\label{fig:mdbexpl}
\end{figure*}

The plan for the first query for the Stock-Level transaction on Cassandra is shown in Figure~\ref{fig:cassexpl}. It can be seen that the filter is successful pushed down to the Multicorn wrapper and then transformed into a filter on the \texttt{key} in the CQL query.
Figure~\ref{fig:mdbexpl} shows the plan for a query in the Order-Status transaction running on the MongoDB data store. It shows how projections and selections are pushed down and combined with the original pipeline query used when defining the foreign table in Figure~\ref{fig:mdbddl}. 
In particular, analysing the plan from the bottom up shows that the optimizer is capable of realizing that the columns referred to in the \texttt{WHERE} statement of the original query as fields of nested documents are also fields in the outer documents, enabling the filter equivalent to the \texttt{WHERE} statement to be pushed-down to be the first operation to be performed, without actually requiring any previous \texttt{\$unwind} operations, thus making this query very efficient. Also, the last \texttt{\$project} operation to be performed (appears first in the listing), selects only the fields that match the specification of the \texttt{SELECT} statement to be returned, thus limiting data transfer to the absolute necessary.

\section{Related work}
\label{sec:related}

Increasing usage of multiple SQL and NoSQL database systems in the same organization, and in particular by the same application, has sparked an interest in tools and techniques that make this simpler. Moreover, it is increasingly important to enable queries across different database systems, combining data from different sources.

Hadoop MapReduce and Hive support multiple file formats and data sources such as NoSQL data stores by means of pluggable input format adapters. These adapters may discover the schema, including partitioning, or impose a user defined one where not available, and then iterate over data. The Hadoop input format interface allows projecting and filtering to be pushed down thus reducing the amount of data moved. The same approach is used by in-memory MapReduce engines such as Spark\,\cite{Armbrust:2015:SSR:2723372.2742797} and Flink, and also by parallel-distributed query engines such as Presto and Impala, to interface with multiple file formats and data sources.
As these systems focus on large scale queries by doing as much as possible of the query computation themselves, such that it can be optimized and parallelized, the ability to exploit particular querying capabilities and operators of the data stores is minimal. Moreover, even if the minimal query latency of the parallel-distributed query engines is much smaller than that of MapReduce, it still isn't adequate to operational workloads.

The alternative is to consider a federated database system with multiple SQL and NoSQL data stores. The SQL/MED standard has support for pushing down parts of queries for remote execution\,\cite{Melton:2002:SSR:601858.601877}. However, there is no single query language that can adequately express queries over such diversity of processing systems, i.e., fully exploit their unique characteristics, and yet these characteristics are precisely the reason these are used [Sto15]. Recent research addressing these challenges has lead to polyglot database systems, also known as polystores.

A possible approach is to start with a high-level SQL query and then compile parts of it to low-level queries that can be pushed to the data stores. The remainder of the query is executed by a traditional query engine, being able to provide additional operators and to combine data from multiple sources. This has been proposed for distributed file systems, using MapRreduce\,\cite{DeWitt:2013:SQP:2463676.2463709}. The same approach is also used by SpliceMachine, that uses the Apache Derby query engine and  translates parts of a query to HBase co-processors.\footnote{\url{https://www.splicemachine.com/product/how-it-works/}} 
More recently, it has been proposed that a full fledged query language for nested\,\cite{Seco:2015:CDM:2815072.2815074} or relational\,\cite{Begoli:2018:ACF:3183713.3190662} can to large extent be translated to a variety of target NoSQL data stores with various data and query models. This uses an optimizer to rewrite parts of the query, driven by cost estimates, while the remainder is executed by a general purpose query engine. The Calcite open source system provides an implementation that is being used, for instance, in Dremio\footnote{\url{https://github.com/dremio/dremio-oss}}, that provides columnar distributed-parallel query execution and materialization. 
F1 Query\,\cite{Samwel:2018:FQD:3229863.3275551}, is a federated query processing platform that executes SQL queries against data stored in different file-based formats as well as different storage systems at Google.
These approaches do not however exploit native query processing capabilities when they expose specialized indexing and query operators. They also don't make it easy to circumvent the translation layer to access native queries. Moreover, those that rely on MapReduce jobs increases end-to-end response time.

An alternative approach that focuses on exploiting the specialized query capabilities of each NoSQL data store is exemplified by BigDawg\,\cite{Duggan:2015:BPS:2814710.2814713,polybases}, that admits that a systems can be mapped to various data models and query languages instead of a single uniform language. This would reduce the conceptual gap between generic languages and target data stores, however, still resulting in a wide variety of concepts to be introduced into the low-code platform. It has also been proposed that intermediate results can be built using a native query on some system and then transferred to a second system for a second computation step, using a different query system\,\cite{LeFevre:2014:MSU:2588555.2588568}, suggesting that materialization can be an important tool in bridging the gap between different data and query models.

The CloudMdsQl approach\,\cite{KVB+15} proposes an extension of SQL where parts of the query can be translated to native queries, but also allowing a query to embed \emph{ad hoc} views defined using native query language snippets. This allows native query capabilities to be fully used when necessary. It also circumvents the need for introspection, as schema is imposed by the query itself. The main advantage is that some of the query can be optimized globally and executed by the top level query engine. For instance, it uses BIND JOIN\,\cite{Haas:1997:OQA:645923.670995} to reduce data transfer across different data stores. This is the closest to our proposal, although we provide schema introspection, at least as a starting point for the developer, and use a standard SQL query engine.


\section{Lessons Learned}
\label{sec:concl}

We discussed the challenges in integrating a variety of NoSQL data stores with the OutSystems low-code platform. This is achieved by a SQL query engine that federates multiple NoSQL sources and complements their functionality, using PostgreSQL with Foreign Data Wrappers as a proof-of-concept implementation. It allowed us to learn some lessons about NoSQL systems and to propose a good trade-off between integration transparency and the ability to take full advantage of each systems' particularities.
 At the same time, it does not constrain the developer, that can override introspection and query generation by using embedded native query language snippets when required. In some cases, this doesn't even constrain the ability for the query engine to globally optimize the query.
The main lessons learned for a \emph{low-code platform provider} are:


\textbf{1. Target an extended relational model.} The relational data model when extended with nested composite data types such as maps and arrays can successfully map the large majority of NoSQL data models with minimal conversion or conceptual overhead. Moreover, when combined with flatten and unflatten operators, the relational query model can actually operate on such data and represent a large share of target query operations. This is very relevant, as it provides a small set of additional concepts that have to be added to the low-code platform or, preferably, none at all as unnesting is done when importing the schema.

\textbf{2. A query engine is needed.} Due to the varying nature of query capabilities in different data sources, a query engine that can perform various computations is necessary to avoid that developers have to constantly mind these differences. This is true even for querying a single source at a time.

For \emph{polyglot developers}, we underline the lessons from CloudMdsQl\,\cite{KVB+15} with one notable exception:

\textbf{3. Basic schema discovery with overrides is needed.} Although CloudMdsQl\,\cite{KVB+15} has shown that it is possible to build a polyglot query engine without schema discovery, by imposing ad-hoc schemas on native queries, it severely restricts its usefulness in the context of a low-code platform. However, after getting started with automatically inferred schema, it is useful to allow the developer to impose additional structure such as composite primary keys in key value stores.

\textbf{4. Embedded scripting is required.} Although many data manipulation operations could be done in SQL at the query engine, embedding snippets of a general purpose scripting language allows direct reuse of existing code and reduces the need for talent to translate them. Together with the ability to override automatic discovery, this is key to ensuring that the developer never hits a wall imposed by the platform.

\textbf{5. Materialized view substitution is desirable.} Although our proof-of-concept implementation does not include it, this is the main feature from the Calcite-based alternative that is missing. The ability to define different native queries as materializations of various sub-queries is the best way to encapsulate alternative access paths encoded in a data-store specific language.

\textbf{6. Combining foreign tables with scripting is surprisingly effective.} Although CloudMdsQl\,\cite{KVB+15} proposed its own query engine, a standard SQL engine with federated query capabilities, when combined with a scripting layer for developing wrappers such as Multicorn, is surprisingly effective in expressing queries and supporting optimizations.

Finally, the main lessons learned for \emph{NoSQL data store providers} are:

\textbf{7. A NoSQL query interface should be targeted at machines, not only at humans.} NoSQL systems such as MongoDB or Elasticsearch, that expose a query model based on an operator pipeline, are very friendly to integration as proposed. In detail, it allows generating native queries from SQL operators or to combine partially hand-written code with generated code. Ironically, systems that expose a simplistic SQL like language that is supposed to be more developer friendly, such as Cassandra, make it harder to integrate as queries in these languages are not as easily composed. 

\textbf{8. Focus on combining query fragments.} It might be tempting to overlook some optimizations that are irrelevant when a human is writing a complete query, e.g., as pushing down \verb|$match| in a MongoDB pipeline. However, these optimizations are fairly easy to achieve and greatly simplify combining partially machine generated queries with developer written queries.



%

\bibliographystyle{splncs04}
\bibliography{paper}  


%
%
%
%

\end{document}